# Mapping nanoscale charge states and phase domains with quantitative hyperspectral coherent diffractive imaging spectroscopy

Allan S. Johnson[1*], Jordi Valls Conesa[1], Luciana Vidas[1], Daniel Perez-Salinas[1], Christian M. Günther[2,‡], Bastian Pfau[3], Kent A. Hallman[4], Richard F. Haglund Jr[4], Stefan Eisebitt[2,3], and Simon Wall[1,5†]

[1]ICFO, BIST, 08860 Castelldefels, (Barcelona), Spain
[2]Institut für Optik und Atomare Physik, Technische Universität Berlin, 10623 Berlin, Germany
[3]Max-Born-Institut, 12489 Berlin, Germany
[4]Department of Physics and Astronomy, Vanderbilt University, Nashville, Tennessee 37235-1807, USA
[5]Department of Physics and Astronomy, Aarhus University, Ny Munkegade 120, 8000 Aarhus C, Denmark
‡Present address: Technische Universität Berlin, Zentraleinrichtung Elektronenmikroskopie (ZELMI), 10623 Berlin, Germany
*allan.johnson@icfo.eu
†simon.wall@phys.au.dk

**The critical properties of functional materials and nanoscale devices often originate from the coexistence of different thermodynamic phases and/or oxidization states, but sample makeup is seldom completely known *a priori*. Coherent diffractive imaging (CDI) provides the spatial resolution needed to observe nanoscale coexistence while returning the full amplitude and phase information of an object, but to date lacks the spectral information necessary for composition identification[1–5]. Here we demonstrate CDI spectroscopy (CDIS), acquiring images of the prototypical quantum material vanadium oxide across the vanadium $L_{2,3}$ and oxygen $K$ X-ray absorption edges with nanometer scale resolution. Using the hyperspectral X-ray image we show coexistence of multiple oxidization states and phases in a single sample and extract the full complex refractive index of $V_2O_5$ and the monoclinic insulating and rutile conducting phases of $VO_2$. These results constrain the role of hidden phases in the insulator-to-metal transition in $VO_2$.**

X-ray spectromicroscopy is a key tool in nanoscience because of its spectral selectivity, which allows unparalleled identification of electronic, chemical, and bond-angle makeup in nanoscale systems[1,2], revealing complex behavior which is otherwise hidden to either spatially or spectrally integrating measurements. Traditional X-ray spectromicroscopy methods are limited by available X-ray optics, restricting their spatial resolution, bandwidth, and sample geometries[1]. To surpass these limits coherent scattering methods like CDI and ptychography have been developed, which do not rely on the quality of the X-ray optics but rather on the coherence of the beam to achieve high spatial resolution[4]. Coherent imaging methods can achieve diffraction-limited resolution while returning the full complex amplitude and phase of the sample. X-ray ptychography has been used to perform limited spectroscopy and measure chemical makeup and charge states[6,7], but requires scanning and suffers from the same sample environment limitations of traditional techniques through its use of zone-plate lenses. Conversely, while CDI does not have such limitations, previous measurements that make use of spectroscopic information have been primarily dichroic and relied on the large contrast of magnetic dichroism effects or the presence of a particular element[8–10], requiring extensive *a priori* knowledge of the sample makeup. Recent broadband CDI measurements require negligible spectral structure in the sample, and return no spectral information whatsoever[11], while proposed hybrid scanning-CDI methods return only coarse spectral structure[12].

A more complex case is that of vanadium oxides like $VO_2$, $V_2O_3$, $V_2O_5$ and $V_3O_7$, which are among the most studied transition metal oxides. In chemistry they form one of the most important families of catalysts[13,14], while in physics they show prototypical metal-to-insulator phase-change behavior in which spin, charge, and lattice degrees of freedom can be coupled[15]. Understanding these complex and spatially inhomogeneous materials requires the ability to identify the local charge state and phase, which requires measurement of the full spectral response of the material and is not possible with spectromicroscopy at a few selected energies alone. Here, using CDIS, we obtain the full spectroscopic and nanometer scale spatial information of a vanadium oxide sample. Using our complete hyperspectral image we show that our sample is a heterogeneous mixture of approximately 80% $VO_2$ and 20% $V_2O_5$. By heating, we observe the phase transition from the monoclinic insulating to rutile conducting phase in the $VO_2$ and extract the full amplitude and phase for three different states simultaneously.

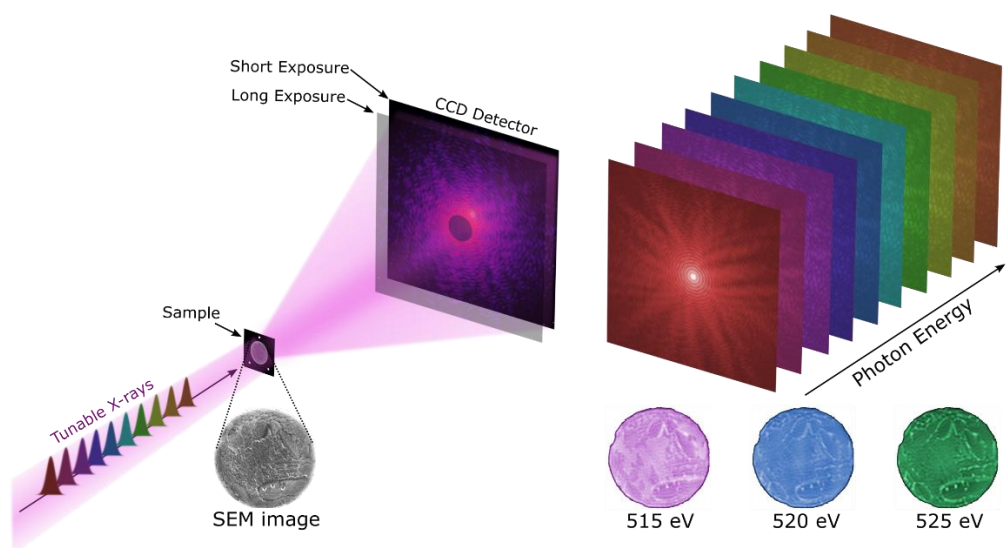

*Figure 1: Experimental setup for coherent diffractive imaging spectroscopy. A tunable synchrotron X-ray radiation source illuminates the sample, and the scattered radiation is collected on a CCD camera. Holographic reference holes in the sample mask provide an absolute phase reference. Long and short exposures are combined to yield a high dynamic range diffraction pattern. Images are recorded at a range of photon energies across the relevant absorption edges; three representative amplitude images are shown.*

A 75 nm thick film of nominally $VO_2$, masked to provide a finite spatial extent, was illuminated with synchrotron radiation and the resulting diffraction pattern recorded on a CCD detector. Reference holes in the mask provide an absolute phase reference allowing us to *quantitatively* extract both components of the complex refractive index $\bar{n}$, not just the relative phase shift between different parts of the sample. A key advantage compared to other X-ray imaging techniques is that no chromatic elements are used, allowing us to scan the X-ray photon energy from 510 eV to 535 eV, without any adjustments in the optical layout between images, to generate a hyperspectral image of the sample with 0.25 eV spectral and 25 nm spatial resolution, totaling 101 spectral images. The required target stability is low, allowing us to easily heat the sample without the need for any active stabilization. Further details on the sample, experimental methods and reconstruction algorithm are given in the Methods section.

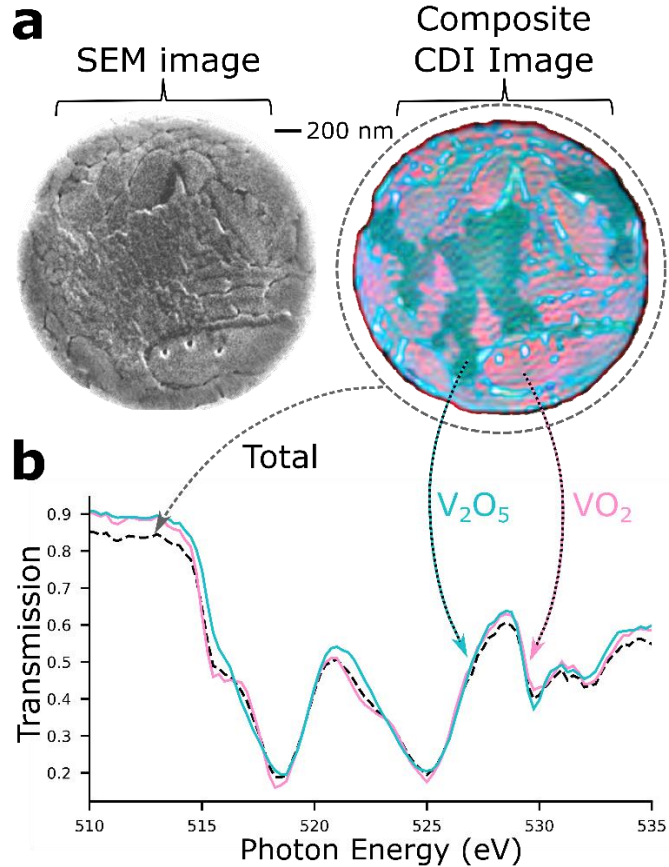

*Figure 2: Hyperspectral imaging of a vanadium thin film.* (a) Scanning electron microscopy image of the sample and comparison to a false-colour composite image of the sample at 320K. Scale bar 200 nm. Two different regions (pink, green) are clearly visible in addition to the topography (blue-white). The composite is formed from a combination of images taken at individual photon images. A total of 101 monochromatic images are taken, allowing us to extract a full transmission spectrum for each pixel in a hyperspectral imaging scheme. (b) Transmission spectra of the pink and green regions of the sample, along with the integrated spectrum as would be observed in absorption spectroscopy (black). The full spectral information allows us to identify the green region as $V_2O_5$ and the pink as monoclinic phase $VO_2$.

The acquired hyperspectral image can be used to generate false-colour images where the different RGB colour channels encode a variety of spectral bands. **Figure 2**a shows a comparison between a scanning electron microscopy (SEM) micrograph and a false-colour amplitude image of the sample at 320 K we compose from eight different frequency bands (see Methods). Thickness changes reflecting the topography of the sample are apparent in all individual images and as blue-white features in the composite image, agreeing very well with the SEM micrograph. Most strikingly however, we can see two major regions of the sample (pink, green), which are not apparent from the SEM image. To identify these regions, we extract the full local X-ray absorption spectra from the hyperspectral image, which are plotted in **Figure 2**b, along with the integrated spectrum of the whole sample. Clear differences appear particularly on the vanadium $L_{2,3}$-edges (≈518 and ≈522 eV). Comparison to literature spectra allows us to identify the pink regions as monoclinic phase (M-phase) $VO_2$ and the green region as $V_2O_5$[16,17], and determine that all crystallites share the same orientation[18]. Despite $V_2O_5$ making up 20% of the sample and the large differences in their spectra, the integrated spectrum is broadly consistent with that of $VO_2$, underlining the necessity of spectroscopic imaging methods in accurately determining local material composition. Interestingly, the boundary between the $V_2O_5$ and $VO_2$ regions does not always correspond to a distinct topographic feature in the SEM or off-resonant CDI images (SI), allowing us to track oxygen diffusion which is not visible to non-spectroscopic imaging.

We next use the coherent nature of CDIS together with our reference structure to examine the metal-to-insulator phase transition of $VO_2$ and extract the complex refractive index of each phase for the first time. As we heat the sample through the $VO_2$ phase transition (≈336 K), regions of the rutile metallic phase (R-phase) begin to nucleate near crystallite edges and other defects, eventually growing to encompass the entire $VO_2$ volume[19] (SI Figure S1). **Figure 3**a shows a false-colour amplitude image of the sample when heated to 336 K, at the center of the transition. Purple regions have grown to encompass roughly 50% of the $VO_2$. An identical structure is seen in the phase, as seen in **Figure 3**b, where – due to the different spectral response in the real and imaginary part of the refractive index – an alternative set of photon energies is used to generate the false colour image (Methods). These new regions are readily identifiable as the R-phase from the absorption spectra, and we can locally extract the complex refractive index of all three constituents ($V_2O_5$, R- and M-phase $VO_2$) from the single hyperspectral data set as shown in **Figure 3**c. The imaginary part of the refractive index $k$ is in excellent agreement with previous absorption measurements for all three constituents, while the real part $n$ provides a new constraint for *ab-initio* models of vanadium oxides[20]. The retrieved $n$ is noisier than $k$ because the phase shift is measured relative to the reference holes, which have low transmission due to their small size, imaged as near single-pixel size. However, as the values at different regions are measured simultaneously and relative to the same reference, this noise cancels when considering the differences in $n$ and $k$. This also ensures the differences are robust with respect to drift during measurement, an advantage of full field-imaging previously noted in conventional X-ray spectromicroscopy[21], but only for absorption or phase independently. The differences in $n$ and $k$ between different constituents are shown in **Figure 3**d and e, and show oscillations characteristic of the resonance structure, with $n$ and $k$ oscillating out of phase. This is consistent with the Kramers–Kronig relation of the resonances, and is a very effective way of identifying the exact energies of overlapping resonances.

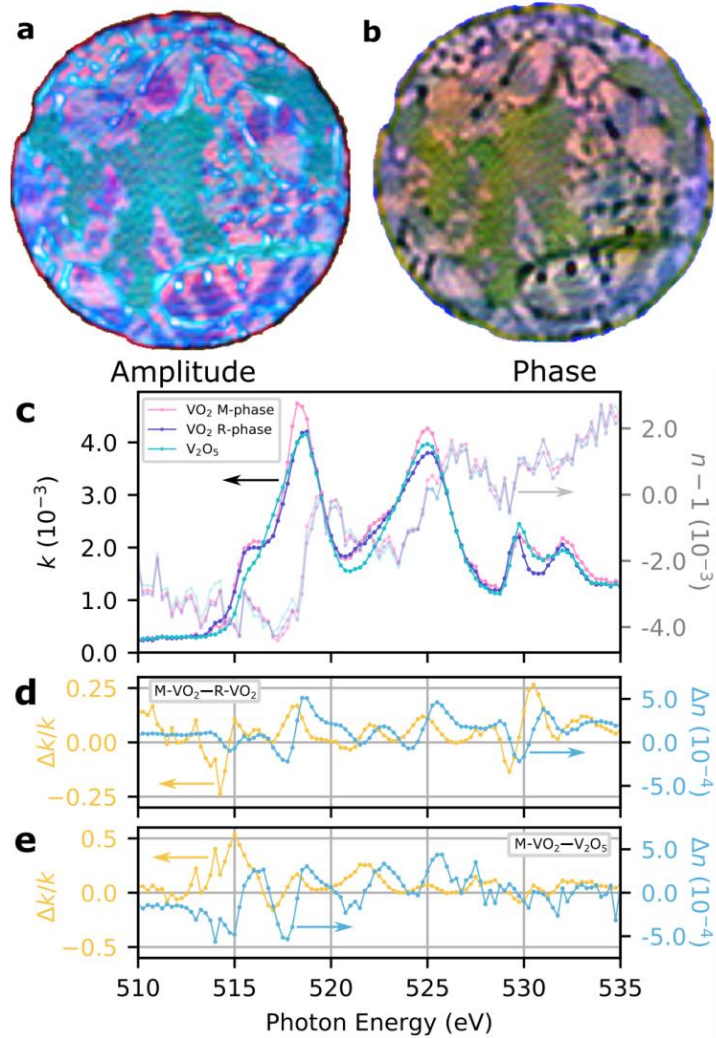

*Figure 3: Fully quantitative amplitude and phase imaging of thermodynamic phase and oxidization state coexistence in vanadium oxide. (a) False-colour amplitude of the VO$_2$/V$_2$O$_5$ sample at 336K. The new purple regions correspond to the R-phase of VO$_2$. Scale bar 200 nm. (b) Corresponding phase image. (c) Extracted real (solid lines) and imaginary (faded lines) parts of the refractive index for V$_2$O$_5$ and the monoclinic and rutile phases of VO$_2$, with colours corresponding to the false-colour amplitude image. (d) Difference between the real and imaginary parts of the refractive index for VO$_2$ R- and M1-phase. (e) Difference between the real and imaginary parts of the refractive index for V$_2$O$_5$ and VO$_2$ M1-phase.*

While these spectra could, in principle, be extracted from the cross correlation of sample and reference with the Fourier Transform Holography (FTH) method, significant low-frequency noise in the reconstruction prevents quantitative spectroscopy (SI section S2). To remove this noise, a beam stop is often used to suppress the central part of the hologram. However, this mixes real and imaginary components making quantitative spectroscopy impossible. Furthermore, amplitude-to-phase coupling via propagation of coherent X-rays can lead to artifacts, which cannot be identified without the absolute division of amplitude and phase (see SI section S3). This artifact led to the erroneous identification of an intermediate state during the phase transformation in previous work based on FTH[19]. The intermediate phase, which was assigned to the M2 phase of VO$_2$, is not observed in the full CDIS data set. Notably the regions of phase coexistence found from the amplitude and phase information are identical, ruling out the presence of strongly strained regions which could induce a secondary transition[22].

In summary, we have extended coherent diffractive imaging to coherent diffractive imaging spectroscopy, generating hyperspectral images of a vanadium oxide thin film spanning 510 eV to 535 eV. We recover the full complex amplitude and phase across the x-ray absorption spectrum, and uniquely identify different oxidization states and phases at temperatures across the $VO_2$ insulator-to-metal phase transition. Our work extends the utility of CDI to cases where the sample makeup is not known *a priori*, opening the door to study of technologically relevant systems where localized states are important, such as in solar cells or catalytic devices[23–25]. Though the current measurements were made with 0.25 eV spectral resolution and 25 nm spatial resolution, state-of-the-art beamlines allow for sub-10 meV spectral resolution if required, while the spatial resolution is set by the scattering geometry and is ultimately diffraction limited. We note that moving between absorption edges separated by hundreds of eV, relevant for many other materials, would require no changes to the setup. In comparison with ptychography or conventional methods we require no scanning of the probe beam or sample as we measure all points of the sample simultaneously, no complex stabilization of the sample position, and no high-numerical-aperture diffractive optics[6,21]. As such, CDIS is fully compatible with time-resolved measurements and with a wide range of sample environments, including high magnetic fields, plasmas, and cryogenic samples, paving the way for multi-dimensional spectroscopic imaging with X-ray probes.

**Methods**

**Sample and Data Acquisition**

A 75 nm thick layer of $VO_2$ was deposited on a $Si_3N_4$ membrane. The opposite surface was coated with a [Cr(5nm)/Au(55nm)]$_{20}$ multilayer (~1.1 μm integrated Au thickness) to block the X-rays. Using a focused-ion beam (FIB), a 2 μm diameter aperture was cut in the Cr/Au multilayer to define the field of view. Three 50 nm – 90 nm diameter reference apertures were FIB-milled through all three layers on a 4.5 μm radius around the central aperture.

The experiments were performed at the ALICE X-ray scattering instrument at the UE52-SGM beamline of the BESSY II synchrotron-radiation source. The sample was heated to between 300 K and 360 K with 0.1 K stability. Linearly polarized X-rays were focused using a long-focal-length Kirkpatrick-Baez-like mirror pair, giving a large focal spot ($\approx$ 60 μm x 120 μm FWHM) approximately 25 cm upstream from the sample itself. Diffraction patterns were acquired with an X-ray CCD placed 40 cm downstream of a sample. Ten long (3.5-4 s) and short (0.03 s) exposures were combined in order to maximize the signal to noise at the low and high amplitude regions prior to reconstruction, with a central beamblock added for the long exposures to avoid camera damage and CCD bleed. The long exposure images were scaled to ensure the integral over the first airy disk ring was the same between both images and combined using with a circular mask with Gaussian edges.

**Image Reconstruction**

The reference holes used to provide an absolute phase reference additionally pin the zeros of the interference fringes, improving reconstruction robustness[9,26,27], but at the cost of increasing the sample size to near typical transverse coherence length of third-generation synchrotrons[28]. This necessitates the use of partial-coherence reconstruction algorithms which return both the complex object and the X-ray coherence function. As such, reconstructions were performed using the partially coherent reconstruction algorithm of Clark *et al.*[29]; 50 iterations of the error reduction algorithm were followed by 100 iterations of the hybrid input-output algorithm[30], ending with another 49 iterations of error reduction. Every 15 iterations the coherence function was updated by 25 iterations of the Lucy-Richardson deconvolution algorithm[31,32]. The use of partially coherent methods was essential to obtain good reconstructions; fully coherent methods introduced significant low frequency noise to the reconstruction. We found no evidence of under-determination for the partial coherent reconstruction of complex objects as described in previous works[26], likely because the reference holes, while only partially coherent, lead to sufficiently clear interference fringes to pin the reconstructed phase. Most critical to obtaining high quality reconstructions was the accurate determination of the mask function, particularly for the hybrid input-output method – a single pixel error in the mask radius introduced significant reconstruction noise. With correct determination of the mask the algorithms were found to converge to the same value regardless of the initial guess. Images were reconstructed at 101 different energies and four different temperatures. By setting the pre-edge transmission to coincide with the atomic scattering factor transmission for a film of 75 nm thickness (90%) we convert the relative transmission values to absolute values. As small drifts in the synchrotron mode over time can adversely affect the spectrum, we used the nominally temperature independent spectrum of the $V_2O_5$ to normalize out these effects.

**False-colour Images**

False-colour images are generated to show contrast for particular phases of $VO_2$, utilizing the different spectral response of the different phases in the real and imaginary part of the refractive index. To that end, the sum- and difference-images at various photon energies are considered in order to take advantage of the full spectral information. For amplitude images the red channel was taken as the sum of images at 514.25 eV and 517 eV with images at 518.25 eV and 515.25 eV subtracted, the green channel images 521 eV and 529.5 eV added with 530.5 eV subtracted, and the blue channel at 517 eV. A DC level of 0.5 was added to the red channel, and the channels were scaled by 1.6, 1, and 1.3, respectively, for plotting. Phase images were composed with the red channel taken as the sum of images at 514.75 eV, 517.75 eV, and 529.75 eV with images at 518.75 eV, 522.75 eV, and 531 eV subtracted, the green channel images 521.5 eV, 524 eV, 525.5 eV, and 528.75 eV added and the blue channel the sum of 515 eV and 517.75 eV with 516.5 eV subtracted. A DC level of 0.5 was subtracted from the red channel, and the channels were scaled by 1, -1, and -1.2, respectively, for plotting. This scheme was chosen instead of a simple mapping of differences between the relative phases to different colour channels for three reasons. One, in order to map different phases to different colours in such a way that the three phases are apparent to viewers with all major types of colour blindness. Two, in order to maximize the spectral information utilized to differentiate different phases, reducing the impact of reconstruction noise or sample contamination, particularly at the oxygen *K*-edge. Third, in order to enable the simultaneous viewing of the sample topography and phase/oxidization state separation, to better understand the relationship or lack thereof between topography and said separation.

**Acknowledgements**

This project has received funding from the European Research Council (ERC) under the European Union's Horizon 2020 research and innovation programme (grant agreement No 758461) and under the Marie Skłodowska-Curie grant agreement No. 754510 (PROBIST), as well as the Ministry of Science, Innovation and Universities (MCIU), State Research Agency (AEI) and European Regional Development Fund (FEDER) PGC2018-097027-B-I00, and was supported by Spanish MINECO (Severo Ochoa grant SEV-2015-0522, SEV2015-0496) as well as Fundació Privada Cellex, and CERCA Programme / Generalitat de Catalunya. L.V. acknowledges financial support by the HZB. Research at Vanderbilt was supported by the United States National Science Foundation EECS-1509740.

**Author Contributions**

S.E. and S.W. conceived the project. K.A.H. and R.F.H. grew the samples, which were processed by C.M.G. subsequently. L.V., D.P.S., C.M.G. and B.P. measured the diffraction patterns. A.S.J., J.V.C. and D.P.S. processed the images and A.S.J. analyzed the data. A.S.J. and S.W. wrote the manuscript with input from all authors.

**Competing interests**

The authors declare no competing financial interests.

# Supplementary Information: Mapping nanoscale charge states and phase domains with quantitative hyperspectral coherent diffractive imaging spectroscopy

Allan S. Johnson[1*], Jordi Valls Conesa[1], Luciana Vidas[1], Daniel Perez-Salinas[1], Christian M. Günther[2,‡], Bastian Pfau[3], Kent A. Hallman[4], Richard F. Haglund Jr[4], Stefan Eisebitt[2,3], and Simon Wall[1,5†]

[1]ICFO, BIST, 08860 Castelldefels, (Barcelona), Spain
[2]Institut für Optik und Atomare Physik, Technische Universität Berlin, 10623 Berlin, Germany
[3]Max-Born-Institut, 12489 Berlin, Germany
[4]Department of Physics and Astronomy, Vanderbilt University, Nashville, Tennessee 37235-1807, USA
[5]Department of Physics and Astronomy, Aarhus University, Ny Munkegade 120, 8000 Aarhus C, Denmark
‡Present address: Technische Universität Berlin, Zentraleinrichtung Elektronenmikroskopie (ZELMI), 10623 Berlin, Germany
*allan.johnson@icfo.eu
†simon.wall@phys.au.dk

**S1 Temperature Dependence of the VO$_2$ domains**

Below the transition temperature only the monoclinic M1 phase of VO$_2$ (pink) and V$_2$O$_5$ (green) are observed. As the sample temperature is increased, metallic domains of R-phase VO$_2$ (dark blue) nucleate and gradually grow to encompass almost all parts of the sample, which previously were in the VO$_2$ M1 phase. At even higher temperature the entire sample switches, though this was not measured here. The temperature dependence of the absorption image (amplitude component) is shown in Figure S1.

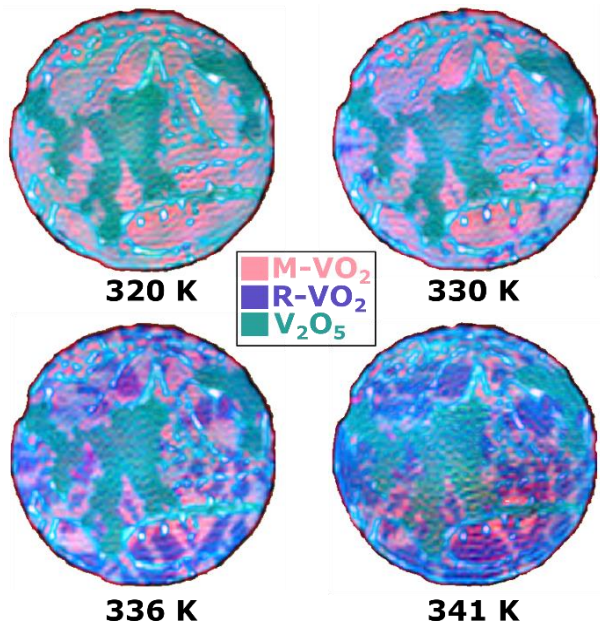

*Figure S1: Temperature dependence of the domain structure in the vanadium oxide thin film. Metallic rutile domains of VO$_2$ nucleate and gradually grow with increasing temperature.*

**S2 Direct Fourier-Transform Holography**

The use of reference holes in our mask allows for a Fourier-Transform Holography analysis[1]. Figure S2 shows the result of this procedure for an acquisition at 336K with 515 eV photons. The central autocorrelation is several orders of magnitude brighter than the images, and it exhibits a large diameter secondary plateau, likely resulting from second harmonic contamination, which overlaps with the images. The overlap with this second order 'halo' prevents meaningful data extraction; in addition to the obvious strong intensity gradient introduced by the overlap, interference between the image and the distribution actually cause some features to switch sign, even relative to their neighbors, or even simply be washed out. The reference holes are placed 4.5 times the radius of the central aperture away from the center and the overall size of the sample approaches the coherence length of 3$^{rd}$ generation synchrotron radiation sources[2], suggesting increasing the spacing between sample aperture and references would have to be traded off against an overall reduced signal due to the necessity of spatial and/or spectral coherence filtering.

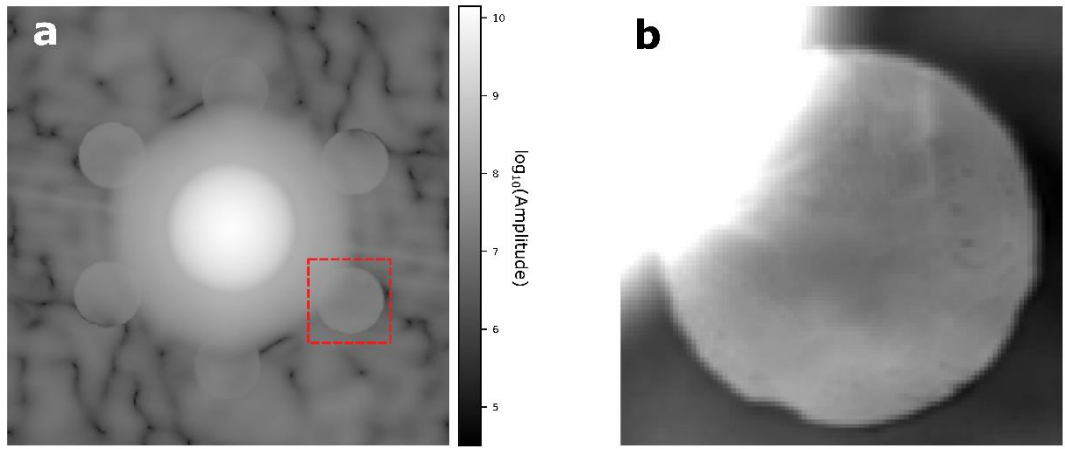

*Figure S2: Fourier Transform Holography analysis. (a) A log10 image of the amplitude channel showing the 6 images (two per reference hole) and the central autocorrelation. The autocorrelation is several orders of magnitude brighter, and a secondary plateau overlaps the sample images. (b) Zoomed in view of the red area in part a, on a linear scale. Features can be discerned, but the autocorrelation overlap prevents quantitative data extraction.*

**S3 Propagation based amplitude to phase coupling artefacts**

All coherent imaging methods naturally return a complex image of the sample at a particular plane in space. For instance, in CDI this plane is set by the plane which best matches the applied spatial constraint (mask)[3]. After reconstruction of the complex real-space image it is then possible to propagate the image to other planes using the phase information. The choice of plane to analyze is most often performed by choosing a feature and propagating until it is in-focus, which is to say when it forms the sharpest image. In non-coherent methods this procedure is generally well defined, but in coherent imaging methods amplitude-to-phase coupling resulting from propagation can lead this choice to no longer be unique as the amplitude and phase components may create a sharp image at different planes which do not correspond to the object plane. For instance if an object were purely absorbing, not imparting a phase shift to transmitted light, propagating the reconstruction until the phase component came into focus would map to the wrong plane and create an apparent but artificial phase component to the sample, as shown in Figure S3. The inverse case is also true, and is the foundation of propagation-based phase-contrast imaging[4]. In order to extract quantitative information as we do with the *n* and *k* of the sample, it is necessary to analyze features in the object plane. However, if the nature of the object being imaged is unknown, this can be challenging.

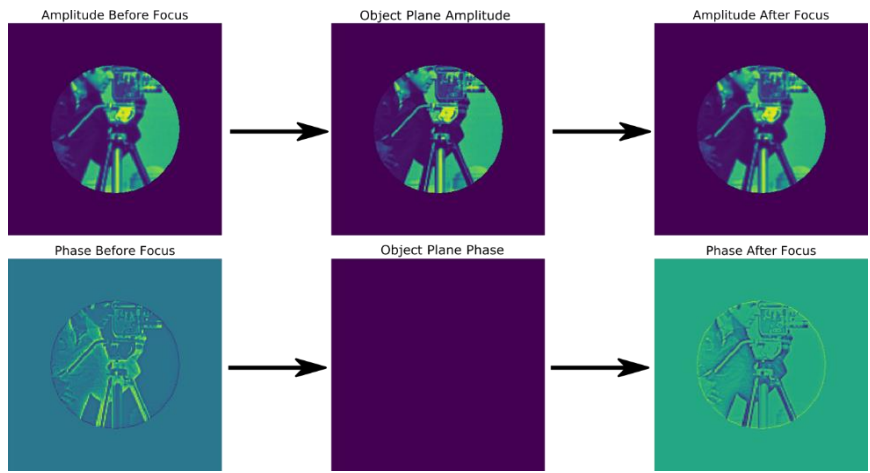

*Figure S3: Effect on propagation of an originally purely amplitude image. Propagation causes features in the amplitude channel to appear in the phase channel. Optimizing on the phase channel would lead to erroneous extraction of the refractive index. The artificial propagation-induced phase changes sign through the real focus at the object plane.*

In CDI the problem can generally be overcome using two pieces of information: one, that the primary contrast mechanism can be discerned *a priori* and propagation performed to bring the correct component into focus, and two, that the artificial propagation-induced amplitude or phase components flip sign as they are propagated through the object plane (Figure S3). These principles allow us to identify the correct sample plane and extract quantitative information. In our particular case real-space images of the sample at 336K were propagated to ensure that the amplitude components in our CDIS image came into focus at peaks of the absorption spectrum, where the contrast is dominantly in the amplitude channel.

As the amplitude-to-phase coupling is a result of the coherent near-field diffraction during propagation, the artifacts emerging from the effect are particularly problematic when using high-pass-filtered diffraction patterns. In many applications of x-ray Fourier-transform holography this filtering results from applying a beamblock to discard the high-intensity DC-Fourier terms from the hologram during data acquisition[1]. While these experiments typically aim for a nanometer-scale image of the sample rather than a quantitative spectral analysis, fringe patterns from filtering and incorrect propagation may confuse the interpretation of the images.

To illustrate this effect in Figure S4 we focus on a particular region of the sample and examine its temperature and propagation dependence. In the optimal focal plane (best object plane estimate), at 336K we see the sample takes on a stripe pattern from the growth of domains. We label three of these regions as P1, P2, and P3. Comparing the amplitude images at 320K and 336K, it is clear that regions P1 and P3 do not change as the temperature is increased, whereas there is a strong change in the region P2 (blue to green), located in between the stripes of P1 and P3. In the phase images these changes are much more subtle. Propagation away from the optimal focal plane causes diffraction from the phase-domain boundaries in the amplitude to create features in the phase image. We now see an almost equally sharp image of the P1-P2-P3 structure in both the amplitude and phase planes. While the same changes with temperature are still apparent in the amplitude channel at focus, in the phase channel it looks as if the regions P1 and P3 change, while P2 does not. When using the real-part of the image reconstruction where amplitude and phase channels are mixed (as in Vidas *et al.*), it is not possible to completely determine which features are real as all regions would be observed to be changing. We believe this effect explains the earlier observations of an intermediate phase based on X-ray holography measurements using high-pass-filtered diffraction patterns only.

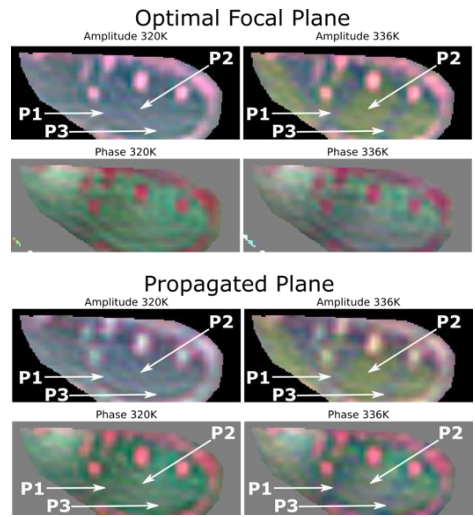

*Figure S4: Propagation induced features in CDI. Top: composite amplitude and phase images at 320K and 336K at the optimal focal plane. As the sample is heated domains grow, splitting the crystal into stripes, 3 of which are labelled P1, P2, and P3. Comparison of the amplitude channel at the two temperatures show that P2 corresponds to the emergent phase. Bottom: Corresponding plots propagating the spectrogram forward in space to the optimal mask plane. Differences now appear in both the amplitude and phase channel. Comparison of the two temperatures show that P2 appears as the emergent phase in the amplitude channel, while in the phase P1 and P3 appear to be emergent. Images are composed with red, green and blue channels corresponding to 518 eV, 530.5 eV, and 529 eV as in Vidas et al.[5]*

The artificial features are stronger when discarding the DC term but keeping the CDI-reconstructed phase, showing that their absence at the focal point is not a result of our reconstruction. When using the correct plane no signature of the intermediate phase is observed, and we see only the direct M1 to R pathway. We note that the ambiguity introduced by amplitude-to-phase coupling is inherent to any coherent imaging method which does not return the absolute amplitude and phase of the image.